\documentclass{aa}
\usepackage[varg]{txfonts}

\usepackage{graphicx}
\usepackage{natbib}
\usepackage{multirow}
\usepackage{amsmath}
\usepackage{amssymb}
\usepackage{url}
\usepackage{color}
\usepackage[normalem]{ulem}

\newcommand{\hd}[0]{HD~189733}
\newcommand{\hr}[0]{HR~7688}

\newcommand{\bs}[2]{^{\beta}\sigma_{#1,#2}}

\newcommand{\bsp}[0]{$\beta\sigma$~procedure}
\newcommand{\bso}[0]{$\beta\sigma$}

\newcommand{\ds}[0]{\texttt{DER\_SNR}}
\newcommand{\dsa}[0]{\texttt{DER\_SNR}~algorithm}

\begin{document}

\title{A posteriori noise estimation in variable data sets}
\subtitle{With applications to spectra and light curves}
\author{S. Czesla \and T. Molle \and 
   J.\,H.\,M.\,M. Schmitt}
\institute{Hamburger Sternwarte, Universit\"at Hamburg, Gojenbergsweg 112, 21029
Hamburg, Germany}
\date{Received ... / Accepted ... }

\abstract{Most physical data sets contain a stochastic
contribution produced by measurement noise or other random sources along
with the signal. Usually, neither the signal nor the noise are accurately known
prior to the measurement so that both have to be estimated a posteriori.
We have studied a procedure 
to estimate the standard deviation of the stochastic
contribution assuming normality and independence, requiring a sufficiently
well-sampled data set to yield reliable results. This procedure is based on estimating the
standard deviation in a sample of weighted sums of arbitrarily sampled data
points and is identical to the so-called \ds\ algorithm for
specific parameter settings. To demonstrate the applicability of our procedure,
we present applications to synthetic data,
high-resolution spectra, and a large sample of space-based light curves
and, finally, give guidelines to apply the procedure in situation not
explicitly considered here to promote its adoption in data analysis.
}


\maketitle

\section{Introduction}
\label{sec:intro}

Measurements of a quantity of interest can almost never be obtained without
contributions of stochastic processes beyond our control. Examples of such
processes in astronomy may be thermal noise from detectors and effects induced
by variable atmospheric conditions in ground-based observations. In other cases,
variability in the measured quantity itself such as stochastic light variations
caused by stellar convection can be among these processes. Whether we
consider this stochastic contribution noise or signal certainly depends on the
adopted point of view.

In the analysis of data, the measurement uncertainty plays an important role.
For instance, the application of $\chi^2$ in tests of goodness of fit
requires knowledge of the uncertainty in
the individual measurements.
If these
cannot be accurately determined a priori as is often the case,
the properties of the noise have to be estimated a posteriori from the
data set itself.

While we may not have accurate knowledge about the actual distribution of the
stochastic noise contribution, normally distributed errors
are often a highly useful assumption, in particular, if only the general magnitude
of the error is known or of interest \citep[e.g.,][Chap.~7]{Jaynes}.
In many cases, including ours, the additional assumption
of uncorrelated errors is made.

Of course, the problem of estimating the noise
from the data is not new, and scientists have been using different approaches to obtain an
estimate of the amplitude of the noise or the signal-to-noise
ratio. Often, some fraction of the data set would be approximated
using for example, a polynomial model via least-squares fitting so that,
subsequently, the distribution of the residuals can be studied to estimate the amplitude of
the noise term. While it remains immaterial whether the model is physically
motivated or not, we usually lack detailed knowledge about both the noise and
the signal. In highly variable data sets it may then be hard to identify
even moderately extended sections, for which we may reasonably presume that they
can be appropriately represented with a model such as a low-order polynomial to
obtain the residuals. Furthermore, every such data set may need individual treatment in
such an analysis. Therefore, techniques with better properties to handle
such data are desirable.

\label{sec:DERSNRalgo}
To obtain reliable noise estimates from spectroscopic data, 
\citet{Stoehr2008} present the \dsa.
Given a set of spectral bins, $f_i$, from which the noise contribution is to be
estimated, \citet{Stoehr2008} propose the following three-step procedure:
First, the median signal, ${\rm med}_i(f_i)$, is determined, where ${\rm
med}_i$ indicates the median taken over the index $i$.
Second, the standard deviation, $\sigma$, of the noise contribution is estimated
from the expression
\begin{equation}
  \sigma = \frac{1.482602}{\sqrt{6}}{\rm med}_i\left(\,| -f_{i-2} + 2f_i -
  f_{i+2} |\,\right) \; ,
\end{equation}
where the factor of $\approx 1.48$ corrects for bias (see
Sect.~\ref{sec:EstimationVarSTD}).
The sum in parenthesis is proportional to the second numerical
derivative. For a sufficiently smooth and well-sampled signal, it
vanishes except for noise contributions and, thus, becomes a measure for the
noise. While \citet{Stoehr2008} found this form the optimal choice
in their applications, they also point out that higher-order derivatives
can be used. Third, the \ds\ noise estimate is computed by dividing the median
flux by the estimated noise.

In this paper, we take up the idea behind the \dsa, viz., to first obtain
a set of numerical derivatives, which we call the $\beta$ sample, from the
data and derive the standard deviation, $\sigma$, of the noise contribution from
it.
In accordance with these steps,
we refer to the more general concept of estimating the noise from numerical
derivatives as the \bsp, which is identical to the \dsa\
for specific parameter settings (see Sect.~\ref{sec:reltodersnr}). We explicitly
treat the case of unevenly sampled data and study
the statistical properties of the resulting noise
estimates. Combining the latter with results obtained from synthetic and real
data sets, we provide suggestions for the general application of the procedure.

\section{The \bsp}
\label{sec:Method}

We assume that we take measurements of some quantity $Y$, whose
value depends on (at least) one other variable $t$, which may be a continuous or discrete
parameter such as time or wavelength. Measurements of $Y$ are taken at specific
instances $t_i$, yielding data points $(t_i, y_i)$, where $y_i$ is the value of
$Y$ measured at $t_i$. We denote the number of data points by
$n_d$.

We further assume that the behavior of $Y$ is given by some (unknown) function
$g(t)$. The only restriction we impose on $g(t)$ is that it may be possible to
expand it into a Taylor series at
any point, $t_0$, so that
\begin{equation}
  g(t) = \sum_{n=0}^{\infty} \left. \frac{1}{n!}\frac{\partial^n
  g(t)}{\partial t^n} \right|_{t=t_0} (t - t_0)^n \; .
  \label{eq:taylor}
\end{equation}
It is sufficient that the expansion can be obtained 
up to a finite order, which will be specified later. 

In addition to $g(t)$, we allow for an additional stochastic contribution to the
data points, which we indicate by $\epsilon_i$. This contribution may,
for example, be attributable to imperfections in the measurements process
or a stochastic process contributing to the data. At any rate, our
measurements will be composed of both terms, so that $y_i =
g(t_i) + \epsilon_i$. We assume the $\epsilon_i$ to be independent
realizations of a Gaussian random variable with mean zero and (unknown)
standard deviation $\sigma_0$. While any random process may produce this
term, we will generally refer to it as noise contribution in the following.
Given these assumptions, we now wish to estimate the
magnitude of the noise term, $\sigma_0$, without specific knowledge about the
underlying process giving rise to $g(t)$. 

\subsection{Constant and slowly varying signals}
\label{sec:slowVariation}
To start, we assume that $g(t)=c$ for any $t$ with some constant $c$.
In this case, we can be sure that the sample of data
points, $y_i$, was drawn from a Gaussian distribution with standard
deviation $\sigma_0$, and
an estimate of the standard deviation in that sample would, thus,
readily provide an estimate of the coveted amplitude of noise, $\sigma_0$.

Now we alleviate the constraints on $g(t)$ and allow
a slowly varying function, by which
we mean that the difference between any two neighboring data points, $y_{i+1}$
and $y_i$, is expected to be dominated by the Gaussian random contribution,
that is, $g(t_{i+1}) - g(t_i) << \sigma_0$. Thus, we effectively postulate
that $g(t)$ can be approximated by a constant between any two
neighboring data points\footnote{Because $g(t)$ is sampled at discrete points,
the evolution between the sampling points remains irrelevant. Nonetheless,
this is a useful assumption.}.
Even in the case of slow variation, however, we may still expect
that the variation in the signal dominates the noise contribution on larger
scales, that is, $g(t_{i+q}) - g(t_{i}) >> \sigma_0$ for sufficiently large
index offset $q > 1$.
Consequently,
the variation may be dominated by the
actual signal, $g(t)$, beyond the sampling scale, and
we may no longer rely on the fact that the
sample of data points is drawn from a Gaussian parent distribution with
standard deviation $\sigma_0$. Thus, an estimate of the standard
deviation in this sample may yield a value arbitrarily far off the
true amplitude of the noise, $\sigma_0$.
While this estimate would always be expected to be larger than $\sigma_0$
given that the $g(t_i)$ and $\epsilon_i$ are uncorrelated, this fact helps
little to mitigate the problem.

Facing this situation, we now construct a sample of
values whose parent distribution is known and entirely determined by the
noise; this set will be dubbed the $\beta$ sample. In particular,
we take advantage of the premise of slow variation.
To obtain the $\beta$ sample, we subdivide
the sample of data points into subsets, each consisting of two consecutive data
points.
For instance, the first subset could comprise the data points $y_0$ and $y_1$,
the second one the points $y_2$ and $y_3$, and so on.
Of course different strategies for subdivisions are conceivable (see
Sect.~\ref{sec:betaSample}).

The idea now is to combine the data points in these
subsets such that the influence of $g(t)$ is minimized.
Under our above assumptions,
this is the case, if we subtract the values of neighboring data points.
Therefore, we proceed by calculating a sample of differences, which we dub the
$\alpha$ sample, for example,
\begin{eqnarray}
  \alpha_m &=& y_{2m+1} - y_{2m} = g(t_{2m+1}) + \epsilon_{2m+1} - g(t_{2m}) -
  \epsilon_{2m} \nonumber \\
  &\approx& \epsilon_{2m+1} - \epsilon_{2m} \,.
  \label{eq:alpha1}
\end{eqnarray}
Here and throughout this work, the index $m$ is used to enumerate the subsets
and subscript the quantities derived from the subsets of data points.
As the $\epsilon_i$ are independent and drawn from a normal distribution, the
$\alpha$ sample will also adhere to a Gaussian distribution with a
standard deviation of $\sqrt{2}\sigma_0$ because the variances of the parent
distributions of the noise terms add.
Consequently, the sample of values
\begin{equation}
  \beta_m = \frac{\alpha_m}{\sqrt{2}} \, ,
\end{equation}
follows a normal distribution with standard deviation $\sigma_0$.
An estimate of the standard deviation in this sample will yield an
estimate of $\sigma_0$ \textit{irrespective} of $g(t)$ as long as it varies
slowly in the above sense.

The procedure used in Eq.~\ref{eq:alpha1} is essentially that of (first-order)
differencing used in time series analyses to remove trends and
produce a stationary time series \citep[e.g.,][]{Shumway}.
In those cases, where the local
approximation of $g(t)$ by a constant becomes insufficient, an approximation by
a higher order polynomial can still lead to a properly distributed
$\beta$ sample.

\subsection{Equidistant sampling}
\label{sec:equallySpaced}
Let $(t_k, y_k)$ with $0 \le k \le M$ be a subsample of $M+1$ data points
extracted from the original data set so that
$t_{k+1} - t_k = \Delta t$. In fact,
data sets with equidistant sampling are not uncommon; examples in astronomy can
be light curves or spectra.
 
If we expand $g(t)$ around the point $t_0$ and
stop the Taylor expansion at order $N$, $g(t)$ is approximated by
\begin{eqnarray}
  g(t_{k}) &=& \sum_{n=0}^{N} \left. \frac{1}{n!}\frac{\partial^n
  g(t)}{\partial t^n} \right|_{t=t_0} (k \cdot \Delta t)^n + O(N+1)
  \nonumber
  \\
  &=&
  T_N(t_{k}) + O(N+1) \;\;\; \mbox{with} \;\;\; t_k = k\,\Delta t \; .
\end{eqnarray}
Therefore, our data points can be approximated by the $N^{\rm th}$-order
polynomial approximation $z_{N,k} = T_N(t_k) + \epsilon_k \approx y_k $, and
we dub $N$ the ``order of approximation''. If $g(t)$ is a polynomial of degree
$\le N$, the equality is exact so that $z_{N,k} = y_k$.

We now wish to linearly
combine $M+1$ approximations, $z_{N,{0 \ldots M}}$, so that all polynomial
contributions cancel, that is,
\begin{equation}
\sum_{k=0}^{M} a_k T_N(t_{k}) = \sum_{k=0}^{M} a_k \sum_{n=0}^{N} \left.
\frac{1}{n!}\frac{\partial^n g(t)}{\partial t^n} \right|_{t=t_0} (k \cdot
\Delta t)^n \overset{!}{=} 0 \; ,
  \label{eq:combine}
\end{equation} 
where both the coefficients $a_k$ and $M$ are yet to be determined. As
Eq.~\ref{eq:combine} is to hold for arbitrary $\Delta t$, it has to hold for all
$0 \le n \le N$ separately.
Combining constant terms into $c_n$, we obtain
\begin{equation}
  \sum_{k=0}^{M} a_k c_n k^n  \overset{!}{=} 0 \;\;\;
   \mbox{and thus} \;\; \sum_{k=0}^M a_k k^n \overset{!}{=} 0 \;\; \forall \;\;  0 \le n \le
   N \; .
\end{equation}
This expression holds for $M = N+1$ and
\begin{equation}
  a_k = (-1)^k \binom{N+1}{k} \; ,
\end{equation}
\citep[e.g.,][Corollary 2]{Ruiz1996}. Thus, the relation $N+2=M+1$ between the
order of approximation, $N$, and the required sample size of data points
is established to make Eq.~\ref{eq:combine} hold.
Along with the $a_k$, any multiple, $\gamma a_k$, 
with $\gamma \in \mathbb{R}$ also solves the equation. As this will
prove irrelevant for our purpose, we set $\gamma = 1$.

From the above, we see that
a set of $N+2$ data point approximations of order $N$, $z_{N,k}$,
can be combined to yield a weighted sum of noise terms
\begin{equation}
  \alpha_m = \sum_{k=0}^{N+1} a_k \, z_{N, k} = \sum_{k=0}^{N+1}
  a_k \, \epsilon_{k} \approx \sum_{k=0}^{N+1} a_k\,y_k \; .
  \label{eq:alpha}
\end{equation}
Again, $m$ is an index, enumerating the subsets
of data points from which the $\alpha_m$ are calculated (cf.,
Sect.~\ref{sec:betaSample}). The last relation in Eq.~\ref{eq:alpha} results
from $y_k \approx z_{N,k}$ and shows that realizations of $\alpha$
can be obtained from measurements, $y_k$, given that the polynomial
approximation is valid so that $y_k - z_{N,k} << \sigma_0$.

In fact, the resulting expression corresponds to that obtained from $N+1$
applications of differencing \citep{Shumway}
and can be understood to be proportional to a numerical
derivative of order $N+1$ as obtained from the forward method of finite
differences \citep[e.g.,][]{Fornberg1988}. When applied to
the $N^{\rm th}$ order polynomial approximation of $g(t)$,
the derivative of order $N+1$ is necessarily zero.

As we assume identically distributed, independent Gaussian noise,
also $\alpha_m$ follows a Gaussian distribution with a
variance, $\sigma_{\alpha, m}^2$, of
\begin{eqnarray}
  \sigma_{\alpha, m}^2 = \sigma_{\alpha}^2 &=& \sum_{k=0}^{N+1} a_k^2 \,
  \sigma_0^2 = \sum_{k=0}^{N+1} \left( (-1)^k \binom{N+1}{k} \right)^2 \,
  \sigma_0^2 \nonumber \\
  &=& \binom{2N+2}{N+1} \; \sigma_0^2 \; .
  \label{eq:sigAlpha}
\end{eqnarray} 
Therefore, the quantity
\begin{equation}
  \beta_m =
  \frac{\alpha_m}{\sqrt{\binom{2N+2}{N+1}}} \; ,
  \label{eq:beta}
\end{equation}
follows a Gaussian distribution with variance $\sigma_0^2$;
the factor $\gamma$ cancels in the
calculation of $\beta$ but not of $\alpha$.
We call a set of
realizations of $\beta$ a $\beta$ sample.
An estimate of the standard deviation of the $\beta$ sample,
is therefore, an estimate of the standard deviation, $\sigma_0$, of
the noise term in the data.

\subsection{The impact of nonequidistant sampling}
\label{sec:neSampling_1}
In Sect.~\ref{sec:equallySpaced} we explicitly demanded equidistant sampling; a
condition which will not
always be fulfilled. We now alleviate this demand
and replace it by the more lenient condition that $t_{k+q} - t_{k} > 0$
holds for any $q > 0$ and study whether
the resulting nonequidistantly sampled measurements, $y_k$, can still be used
in the construction of the $\beta$ sample although equidistant sampling was
explicitly assumed in Sect.~\ref{sec:equallySpaced}.

As we still demand strictly and monotonically increasing $t_k$,
there is a function $E$ so that $E(e_k) = t_k$, where the $e_k$, again, denote
equidistant points. Therefore, sampling $g(t)$ at the points $t_k$ is equivalent
to sampling the function $g_E(t) = (g \circ E)(t)$ equidistantly at the points
$e_k$, specifically,
\begin{equation}
  y_k = g(t_k) + \epsilon_k = g_E(e_k) + \epsilon_k \; .
\end{equation}
Consequently, the method remains applicable. However, the function to be locally
expanded into a Taylor series would become $g_E$ instead of $g$ itself.
Whether this proves disadvantageous for the applicability of the method depends
on the specifics of the functions $g$ and $E$.

In the special case that $E$ corresponds to the inverse function of $g$,
we obtain $g_E(t) = (g \circ E)(t) = (g \circ g^{-1})(t) = t$, which yields a
trivial Taylor expansion. Such a case appears, however, rather
contrived in view of realistic data sets.

In the more general case that $g$ corresponds to a polynomial $G$ of degree
$n_p$ and $E$ to a polynomial $H$ of degree $m_p$, the composition $g_E(t) =
(G\circ H)(t)$ becomes a polynomial $G_H$ of degree $m_p\times n_p$.
Therefore, if $G$ is a constant (i.e., $n_p=0$) any transformation of
the sampling remains irrelevant because $G_H$ remains a constant. Similarly, any linear
scaling of the sampling axis (i.e., $m_p=1$) leads to identical degrees for $G$
and $G_H$, which remains without effect on the derivation of the noise according
to Sect.~\ref{sec:equallySpaced}. However, a more complex transformation of the
sampling axis leads to an increase in the degree of the polynomial
$G_H$. Therefore, a higher order of
approximation, $N$, may be required to obtain acceptable results.

Missing data points, even in a series of otherwise equidistantly
sampled data, pose a particular problem because the
``discontinuities'' introduced by them often require high polynomial degrees
for $H$ so that the resulting function, $G_H$, also becomes a high order
polynomial even if $G$ itself is a low order polynomial with degree $\ge 1$.
If such gaps in the data are sparse, the corresponding realizations of $\beta$
may simply be eliminated from the $\beta$ sample, and the data set can
still be treated as equidistant.
If the gaps are numerous or the sampling is generally
heterogeneous, the application of the procedure introduced in Sect.~\ref{sec:equallySpaced}
becomes problematic, and a more explicit treatment of the sampling is required.

\subsection{Treatment of arbitrary sampling}
In the general case of nonequidistant sampling, the derivation of the
coefficients $a_k$ has to be adapted. Again, we assume that $t_{k+q} - t_{k}
> 0$ holds for any $q > 0$.
In this case,
Eq.~\ref{eq:combine} will assume the more general form
\begin{equation}
\sum_{k=0}^{M} a_k T_N(t_{k}) = \sum_{k=0}^{M} a_k \sum_{n=0}^{N} \left.
\frac{1}{n!}\frac{\partial^n g(t)}{\partial t^n} \right|_{t=t_0} (t_{k} - t_0
)^n \overset{!}{=} 0 \; .
  \label{eq:combine_2}
\end{equation}
In Sect.~\ref{sec:Method}, we found that in the case of equidistant sampling,
sets of coefficients $a_k$, solving the
equation, can be determined for $M=N+1$. As equidistant sampling is a special
case of the more general situation considered here and, again, realizing that
the equation must hold for any power of $t_{k} - t_0 = t_{k,0}$ separately,
we arrive at the following linear, homogeneous system of
equations for the coefficients $a_k$
\begin{equation}
\begin{pmatrix}
1 & 1 & \ldots & 1 \\
0 & t_{1,0} & \ldots & t_{N+1,0} \\
 & & \vdots & \\
0 & t_{1,0}^N & \ldots & t_{N+1,0}^N \\
\end{pmatrix}
\begin{pmatrix}
a_0 \\
a_1 \\
\vdots \\
a_{N+1} \\
\end{pmatrix}
=
\mathbf{T}\cdot\vec{a} =
\begin{pmatrix}
0 \\
0 \\
\vdots \\
0 \\
\end{pmatrix} \; .
\label{eq:eqsys}
\end{equation}
The solution of this system of $N+1$ equations for $N+2$ coefficients is a
vector subspace whose dimension is given by $N+2 - \mbox{rank}(\mathbf{T})$.
Because $\mathbf{T}$ is a so-called Vandermonde matrix and the $t_k$ are
pairwise distinct, $\mbox{rank}(\mathbf{T})$ is $N+1$, and the dimension of the
solution space is one.

By making the (arbitrary) choice $a_0 = 1$, we can rewrite Eq.~\ref{eq:eqsys}
into the system
\begin{equation}
\begin{pmatrix}
 1 & \ldots & 1 \\
 t_{1,0} & \ldots & t_{N+1,0} \\
 & \vdots & \\
 t_{1,0}^N & \ldots & t_{N+1,0}^N \\
\end{pmatrix}
\begin{pmatrix}
a_1 \\
\vdots \\
a_{N+1} \\
\end{pmatrix}
=
\mathbf{T'}\vec{a'}
=
-
\begin{pmatrix}
1\\
0\\
\vdots\\
0\\
\end{pmatrix}
=
-
\vec{t}_0
\; ,
\label{eq:eqsys2}
\end{equation}
where $\mathbf{T'}$ is a quadratic matrix. The solution of this system of
equations can be obtained by inverting the matrix $\mathbf{T'}$ so that
\begin{equation}
  \vec{a'} = -\mathbf{T'^{-1}} \vec{t}_0 \; ,
\end{equation}
where only one column of $\mathbf{T'^{-1}}$ is actually required.
A solution to Eq.~\ref{eq:eqsys} reads
\begin{equation}
  \vec{a} =
  \begin{pmatrix}
  1 \\
  a_1 \\
  \vdots \\
  a_{N+1} \\
  \end{pmatrix} \; .
\end{equation}
As in the case of equidistant sampling, any multiple, $\gamma \vec{a}$, solves
the equation, reflecting our freedom in the choice of $a_0$ and the
dimensionality of the solution vector subspace.
Using the coefficients $a_k$, realizations of $\beta$
can be obtained analogous to Eq.~\ref{eq:beta}
\begin{equation}
  \beta_m = \frac{\sum_{k=0}^{N+1} a_k \, z_{N, t_{k}}
  }{\sqrt{\sum_{k=0}^{N+1} a_k^2} } =
  \frac{\sum_{k=0}^{N+1} a_k \, \epsilon_{k}}{ \sqrt{\sum_{k=0}^{N+1} a_k^2 }}
  \approx \frac{\sum_{k=0}^{N+1} a_k y_k}{ \sqrt{\sum_{k=0}^{N+1}
  a_k^2 }} \; .
\end{equation}
Again, these will adhere to a Gaussian distribution with standard deviation
$\sigma_0$ if the polynomial approximation is valid. Clearly, the equidistant
case presented in Sect.~\ref{sec:equallySpaced} is a special case of this more
general approach. The obvious, yet only
technical, advantage of equidistant sampling is that the coefficients, $a_k$,
need to be obtained only once, while they could be different for
all subsamples in the case of arbitrary sampling.  

\subsection{Construction of the $\beta$ sample}
\label{sec:betaSample}

To estimate the standard deviation of the error term, $\sigma_0$, from data,
we need to choose an order of approximation, $N$, and obtain
realizations of $\beta$ by calculating weighted sums of subsets of
$N+2$ data points. The question to be addressed here is how these subsets are
selected.

As the order of data points is essential in the polynomial approximation,
we focus on subsets of consecutive data points in the construction of the
$\beta$ sample. For instance, $\{y_1, y_2, y_3\}$ is a conceivable subset.
To account for possible correlation between consecutive data points,
we allow larger spacing between the data points by introducing a jump parameter,
$j$, which can take any positive integer. Any subset contains
the data points $y_{i_m+k\,j}$ with $0 \le k < N+2$ and $i_m$ denoting the
starting index of the $m^{\rm th}$ subset. Thus, for instance $\{y_1, y_3,
y_5\}$ is an admissible subset for a jump parameter of two. 
Depending on how the subsets are distributed (i.e., the values of the $i_m$ are
chosen) and whether they overlap, $\beta$ samples with different statistical
properties can be constructed.

\subsubsection{Correlation in the $\beta$ sample}
\label{sec:corrBeta}

Let $K$ and $J$ be two sets of $N+2$ indices, each defining a subset used in
the construction of the $\beta$ sample.
While any index must occur only once in each set, $K$ and $J$ may contain
common indices; take as an example $K_1=\{5,6,7\}$ and $J_1=\{7,8,9\}$.
Based on $K$ and $J$, realizations of
$\beta$ can be obtained
\begin{equation}
  \beta_K = \sqrt{f^{-1}}\, \sum_{k=0}^{N+1} a_k \epsilon_{K_k} \;\;\;
  \mbox{and} \;\;\; \beta_J = \sqrt{f^{-1}}\, \sum_{j=0}^{N+1} a_j \epsilon_{J_j} \; ,
\end{equation}
where $K_k$ denotes the $k^{\rm th}$ element in $K$ and $f=\sum a_k^2$. We also
assume that the same coefficients $a_k$ apply and the polynomial approximation
holds.

The covariance, $\mathrm{Cov}$, between $\beta_K$ and $\beta_J$ is given by
\begin{equation}
   \mathrm{Cov}\left(\beta_K, \beta_J\right) = \mathrm{E}\left[\beta_K \times
   \beta_J \right] = \ \mathrm{E}\left[\frac{1}{f} \sum_{k=0}^{N+1} \sum_{j=0}^{N+1} a_k a_j
   \epsilon_{J_j} \epsilon_{K_k} \right] \, ,
\end{equation}
where $\mathrm{E}$ denotes the expected value. If $K$ and $J$ have no
elements in common, the covariance is zero. If, however, $K$ and $J$ have a
total of $q$ elements in common and $q_{K,i}$ and $q_{J,i}$ denote the positions
of the $i^{\rm th}$ such common index in $K$ and $J$, the covariance becomes
\begin{equation}
   \mathrm{Cov}(\beta_K, \beta_J) =
   \mathrm{E}\left[\frac{1}{f} \sum_{k=0}^{N+1} \sum_{j=0}^{N+1} a_k a_j
   \epsilon_{J_j} \epsilon_{K_k} \right] = \frac{\sigma_0^2}{f} \sum_{i=1}^{q}
   a_{q_{K,i}} a_{q_{J,i}} \; .
\end{equation}
The previously defined sets $K_1$ and $J_1$ have the index seven in common with
$q_{K,1}=2$ and $q_{J,1}=0$, which yields a covariance of
$\mathrm{Cov}\left(\beta_{J_1}, \beta_{J_1} \right) = a_2a_0\sigma_0^2 f^{-1}$.

While completely distinct index sets $K$ and $J$ guarantee independent
realizations of $\beta$, we note that independent realizations of $\beta$ can
also be constructed with different index sets. For example, the sets
$K=\{1,2,3,4\}$ and $J=\{3,1,4,2\}$, that is, a reordering of data points,
yield independent realizations of $\beta$ for equidistant sampling. 
However, as the actual realizations of $\beta$ are constructed from the data,
only possibilities, which maintain the order of indices are of
interest in this context. 

A case of high practical importance is that of ``shifted sets'' by which we
mean that the indices in $J$ are computed from those in $K$ by adding an integer
$s$ (e.g., $K=\{5,6,7\}$ and $J=\{6,7,8\}$ with $s=1$); this is equivalent to
shifting the starting point, $i_m$, by $s$ data points. For $0 < s < N+2$, the
resulting covariance and correlation, $\rho$, read
\begin{equation}
  \mathrm{Cov}(\beta_K, \beta_J) =
   \frac{\sigma_0^2}{f} \sum_{k=s}^{N+1} a_k a_{k-s} =  \sigma_0^2 \,
   \rho(\beta_K, \beta_J) \; .
\end{equation}
In the case of evenly sampled data,
the $a_k$ have alternating sign, so that also all products $a_k a_{k-s}$ have
the same sign. Therefore, the covariance cannot be zero for any $s < N+2$.
The sign of the covariance is positive for even $s$ and negative for odd $s$.
Consequently, uncorrelated realizations of $\beta$ cannot be obtained from
overlapping shifted subsets of data points.

\subsubsection{Obtaining independent realizations of $\beta$}
Realizations of $\beta$ are independent if any data point is
used only in one realization (Sect.~\ref{sec:corrBeta}).
For a jump parameter of one (i.e., consecutive data points),
the most straight-forward approach to construct such a $\beta$ sample is
to divide the data set into subsets of $N+2$ consecutive data points so that the
first subset comprises the points $y_0 \ldots y_{N+1}$, the second one the
points $y_{N+2} \ldots y_{2\times(N+2) - 1}$, and so on. Incomplete sets with
less than $N+2$ points are disregarded.

If a larger jump parameter
is specified, we define consecutive chunks of $(N+2)\times
j$ points and subdivide them into $j$ subsets by collecting all points for which
$i~\mbox{mod}~j = l$, where $i$ is the index of the point and $0 \le l < j$.
Table~\ref{tab:betaSamDemo}
demonstrates a number of subdivisions following this procedure.
The size of the resulting $\beta$ sample will approximately be given by $n_d\,
(N+2)^{-1}$ not accounting for potentially neglected points in incomplete
subsets.

\begin{table}[h]
\centering
\caption{Demonstration of subsample selection for independent $\beta$ samples.
The assignment to subsamples x, y, z, for various combinations of order of
approximation, $N$, and jump parameter, $j$, is shown, based on a hypothetical
sample of seven data points with indices zero through six; a $-$ indicates data
points not assigned to any subset.
\label{tab:betaSamDemo}}
\begin{tabular}{l l | c c c c c c c } \hline\hline
N & j & 0&1&2&3&4&5&6 \\ \hline
0 & 1 & x&x&y&y&z&z&$-$ \\
0 & 2 & x&y&x&y&z&$-$&z \\
0 & 3 & x&y&z&x&y&z&$-$ \\
1 & 1 & x&x&x&y&y&y&$-$ \\
1 & 2 & x&y&x&y&x&y&$-$ \\ \hline
\end{tabular}
\end{table}

\subsubsection{Obtaining correlated realizations of $\beta$ by shifting}
A $\beta$ sample can be constructed by
assigning the data points $y_{i_0+k\,j}$ ($0 \le k < N+2$) to the first subset
and then increase $i_0$ from zero to  $n_d - j\times (N+1)-1$
in steps of one.
The effect is that of shifting the subsets across the
data. In this way, all data points will occur in the $\beta$ sample at least
once.

The size of the $\beta$ sample obtainable in this fashion is $n_d -
j\times (N+1)$, which hardly depends on the order of approximation and can be
substantially larger than the sample of independent realizations of $\beta$. However,
these realization are correlated (Sect.~\ref{sec:corrBeta}).

\subsection{Estimating the variance and standard deviation}
\label{sec:EstimationVarSTD}

The procedure used to estimate the standard deviation of the $\beta$
sample remains immaterial for the concept of the method and can be chosen to
fit the purpose best.
The usual variance estimators\footnote{We use hats to indicate the
estimator as opposed to a specific estimate.} are
\begin{eqnarray}
  \hat{s}^2(B) &=& \frac{1}{n-1}\sum_{i=1}^n (\beta_i - \overline{\beta})^2
  \;\;\; \mbox{and} \;\;\; \\
  \hat{s}^2_E(B) &=& \frac{1}{n}\sum_{i=1}^n (\beta_i -
  \mathrm{E}[\beta]) ^2 \; ,
\end{eqnarray}
where $\mathrm{E}[\beta]$ denotes the expectation value of $\beta$ and $n$ is
the sample size.
Using $\hat{s}^2$, the expectation value is estimated from the sample by the mean,
$\overline{\beta}$.
While $\mathrm{E}[\beta]$ is zero by construction, in working with real data, the
assumptions leading to this result may not be completely fulfilled, so
that both can be useful in the analysis.
For independent, Gaussian $\beta$ samples, both $\hat{s}^2$ and $\hat{s}^2_E$
are unbiased.
In the case of correlated samples however, only $\hat{s}^2_E$ remains so, while
$\hat{s}^2$ generally becomes biased \citep{Bayley1946}. 

The variance, $\mathrm{V}$, of the estimator $\hat{s}_E^2$ reads
\citep{Bayley1946}
\begin{equation}
  V\left(\hat{s}_E^2\right) = \frac{2\sigma^4}{n} + \frac{4\sigma^4}{n^2}
  \sum_{j=1}^{N+2} (n-j) \rho^2_j \approx \frac{2\sigma^4}{n} \left(1 +
  2\sum_{j=1}^{N+2} \rho_j^2 \right) \, ,
\end{equation}
where $\rho_j$ is the correlation at an offset of $j$ in the $\beta$ sample, and
the approximation holds for $N+2 << n$.

The square root of the variance estimator, $\hat{s}_E(B)$, is an
estimator of the standard deviation. We note however, that it is not
unbiased. For independent Gaussian samples
the expectation value of $\hat{s}_E$ becomes
\begin{equation}
  \mathrm{E}\left[\hat{s}_E(B)\right] = \sigma_0
  \sqrt{\frac{2}{n}}\,\frac{\Gamma\left(\frac{n}{2}\right)}{\Gamma\left(\frac{n-1}{2}\right)}
  \approx \sigma_0 \left(1 - \frac{3}{4n} \right) \; ,
\end{equation} 
where $\Gamma$ denotes the gamma function \citep[see][Chapt.~VII]{Forbes2011,
Kenney}. The bias decreases as the inverse of the
number of independent samples, which is, however, not trivially obtained in the
case of correlated samples \citep{Bayley1946}.

In many practical cases, expectation biases are a minor issue because, first,
they tend to decrease with sample size for Gaussian samples, and second,
the assumption of independent Gaussian errors will generally only hold
approximately.
We will use $s_E$ as an estimate of the
standard deviation of the error term and approximate the standard deviation,
$\sigma_{s_E}$, of the estimate by error propagation based on the variance
of $s_E^2$, that is,
\begin{equation}
\sigma_{s_E} = \sqrt{\frac{V\left(s_E^2\right)}{4 s_E^2}} \; .
\end{equation}

Frequently, the data contain
grossly misplaced data points in
stark contrast with the distribution of the remaining ensemble of points, which
have to be dealt with. Such
points are usually referred to as outliers. Because the breakdown point of the
estimator $\hat{s}_E^2$ is zero \citep{Hampel74}, a single such outlier can be
sufficient to spoil the estimate obtained from it and, thus,
render the result useless. In such cases, a more robust estimator is desirable.

Among the most robust estimators of the standard deviation is the
``median absolute deviation about the median'' (MAD). The estimator is given
by
\begin{eqnarray}
  \hat{s}_{\rm M}(B) &=& k \times \mbox{med}_i\left( \left| \, \beta_i -
  \mbox{med}_j(\beta_j) \, \right| \right) \;\;\; \mbox{and} \\
  \hat{s}_{\rm ME}(B) &=& k \times \mbox{med}_i\left( \left| \, \beta_i \, \right| \right)
  \; ,
\end{eqnarray}
where the constant
$k$ is about $1.4826$ \citep{Hampel74, Rousseeuw}. Again, we can take advantage
of the known median of the $\beta$ sample. While $\hat{s}_{\rm M}(B)$ is a
robust estimator for the standard deviation with the maximal breakdown point of $0.5$,
its asymptotic efficiency in the case of normally distributed samples is only
$37$\% \citep{Hampel74, Rousseeuw}. The price for a more robust estimate is
therefore efficiency. The estimator $\hat{s}_{\rm M}(B)$ is consistent
\citep{Rousseeuw} and thus, asymptotically unbiased.
Besides $\hat{s}_{\rm M}(B)$, a number of alternative, robust
estimators with partly higher efficiency have been given in the literature
\citep{Rousseeuw}, either of which could be used to estimate the standard
deviation of the $\beta$ sample.
In the following, we will concentrate on $\hat{s}_E(B)$ and $\hat{s}_{\rm
ME}(B)$ and refer to these estimators as the minimum variance
(MV) and the robust estimator.

\subsection{Relation to the \dsa}
\label{sec:reltodersnr}
At this point, we have assembled all pieces to see the relation to the
\dsa\ presented by \citet{Stoehr2008}. If the
order of approximation is one, the jump parameter is two, a
correlated $\beta$ sample is constructed by the shifting technique, and
$\hat{s}_{\rm ME}$ is used to estimate the standard deviation from it, the
\bsp\ becomes identical to the \ds\ algorithm presented by \citet{Stoehr2008},
given that the data are equidistantly sampled (see Sect.~\ref{sec:DERSNRalgo}).

\subsection{An upper limit on the amplitude of noise}
\label{sec:upperNoiseLim}

Even if the noise distribution is normal and an unbiased estimator for the
standard deviation is used, the expectation value of the noise estimate is only
identical to $\sigma_0$ as long as $y_i = z_{N,i}$ or, equivalently,
the polynomial approximation is exact $T_N(t_i) = g(t_i)$.
If this is not the case, so that $y_i - T_{N}(t_i) = 
\epsilon_i + r_{N,i}$ with some non-negligible remainder $r_{N,i}$, the noise
estimate may no longer remain unbiased.
Even in the case of $\epsilon_i = 0$, the expectation value
will then generally be larger than zero
because the $\epsilon_i$ are independent of
the $r_{N,i}$ and their variances will add.
While the expectation value of the estimator therefore, cannot be
smaller than $\sigma_0$, we note that this may still be the case for individual
estimates.

\subsection{Relative efficiency of the estimator}

The variance of the estimator $\hat{s}^2_E$ and thus, its efficiency
depend on the size of the $\beta$ sample and the correlation between its
elements.
If an independent $\beta$ sample is constructed, its sample size is
approximately given by $n_d\times(N+2)^{-1}$.
For the lowest order of approximation (i.e., zero), the variance of $s_E^2$
obtained from this sample reads $4\sigma_0^4 n_d^{-1}$. If $s_E^2$
is calculated from the $\beta$ sample obtained by the shifting procedure, the
sample size is approximately $n_d$, and the variance reads $2\sigma_0^4
n_d^{-1} (1+0.5) = 3\sigma_0^4 n_d^{-1}$. We note, however, that this
result only applies for equidistant sampling.

\begin{figure}
  \includegraphics[width=0.49\textwidth]{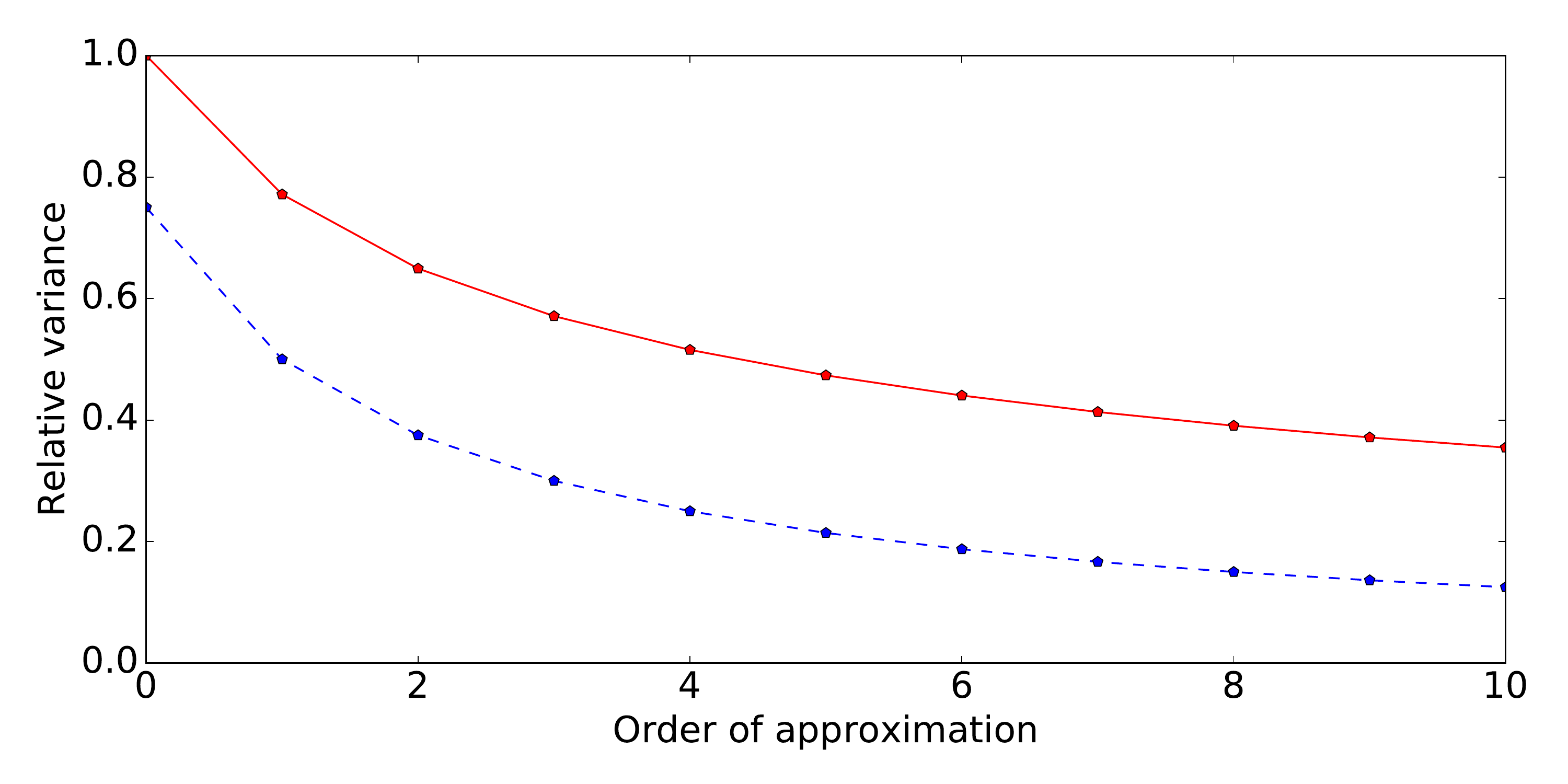}
  \caption{Minimum variance of $3\sigma_0^4 n_d^{-1}$ for $\hat{s}_E^2$
  divided by variance obtained for various orders of approximation for
  independent (dashed blue) and correlated $\beta$ samples (solid red).
  \label{fig:relVar}}
\end{figure}

When orders of approximation $N_1$ and $N_2$ are used with an independent
$\beta$ sample, the ratio of the variances, $V$, of the estimators
$\hat{s}_E^2(N_{1,2})$ reads
\begin{equation}
  \frac{V\left(\hat{s}_E^2(N_1)\right)}{V\left(\hat{s}_E^2(N_2) \right)} =
  \frac{N_1+2}{N_2+2} \; .
\end{equation}
If a correlated $\beta$ sample is constructed using the shifting algorithm, its
size hardly depends on the order of approximation given that $n_d >> N+2$. Due to the
correlation, this does not translate into identical efficiency, however. In
Fig.~\ref{fig:relVar}, we show the variance of $\hat{s}_E^2$ for various orders
of approximation, relative to $3\sigma_0^4 n_d^{-1}$.

Higher orders of approximation always worsen the relative asymptotic
efficiency of the estimator due to the decrease in sample size and the impact of
correlation. The results obtained from the correlated $\beta$ sample are
always superior in terms of efficiency for the same order of approximation.
The correlation structure considered here, however, only applies for equidistant
sampling of the data.

\subsection{Correlation between estimates of different order}
\label{sec:estiCorr}
Given a sample of independent realizations of a Gaussian random variable, the
conditions for the \bsp\ to yield reliable estimates are
fulfilled for all orders of approximation.
Here, we demonstrate that
estimates obtained with different orders of approximation are highly
correlated.

To this end, we obtain a sample of $1000$ independently
sampled Gaussian random numbers and estimate their variance by using $\beta$
samples constructed using the shifting technique and orders of approximation
between zero and four. Repeating this experiment $5000$ times, we estimate the
correlation of the thus derived noise variance estimates and
summarize the results in Table~\ref{tab:orderCorr}; for
clarity we omit the symmetric part of the table. Clearly, high degrees of correlation are
ubiquitous.

\begin{table}
\caption{Correlation of the noise variance estimates.
\label{tab:orderCorr}}
\begin{tabular}{l l l l l l} \hline\hline
  N & 0 & 1 & 2 & 3 & 4 \\ \hline
  0 &  1.000 &  0.977 &  0.943 &  0.912 &  0.884 \\
  1 &  &  1.000 &  0.991 &  0.975 &  0.956 \\
  2 &  &  &  1.000 &  0.995 &  0.985 \\
  3 &  &  &  &  1.000 &  0.997 \\
  4 &  &  &  &  &  1.000 \\
  \hline
\end{tabular}
\end{table}

Practically, this implies that
estimates based on different (and
especially subsequent) orders of approximation differ less than suggested
by their nominal variance, if the order of approximation is
sufficient to account for the systematic variation in the data. As the estimates are based on the
same input data, this behavior may of course be expected.
Naturally, the correlation depends on the
construction of the $\beta$ sample. For example, if an independent $\beta$ sample is
constructed, the correlation is still significant but lower with correlation
coefficients around $0.5$.

\section{Python implementation of \bsp}
Along with this presentation, we provide a set of \texttt{Python} routines
implementing the \bsp\ as outlined here. The code is
available both as a stand-alone module and as part of our PyAstronomy package
available via the github
platform\footnote{\url{https://github.com/sczesla/PyAstronomy}}.

The code runs under both Python~2.7 and the 3.x series and
comes with documentation and examples of application.
It provides algorithms to derive independent and correlated $\beta$ samples
based on both equidistantly sampled and arbitrarily sampled data sets.
Our implementation is distributed under the MIT
license\footnote{https://opensource.org/licenses/MIT}
inviting use and modification by all interested
parties.
The routines are also used in the
applications presented in the following Sect.~\ref{sec:Application}.

\section{Application to real and synthetic data}
\label{sec:Application}
In the following we apply the \bsp\ to estimate the amplitude
of noise in synthetic data, high-resolution echelle spectra, and a sample of
space-based CoRoT light curves to demonstrate its applicability
in real-world scenarios.

\subsection{Application to synthetic data}
\label{sec:sineApp}
We start by determining the noise in a series of synthetic data
sets generated from the function
\begin{equation}
  g(t) = \sin\left(2\pi t \, P^{-1} \right) \; ,
\end{equation}
sampled at a total of $1000$ equidistant points given by $t_i = 0.1 \times i$
($0 \le i < 1000$) so that $\Delta t=0.1$. In our calculations, we use a number
of different periods, $P$,
to demonstrate the effect of different sampling
rates. Specifically, a single oscillation is sampled by $P \Delta t^{-1}$ data
points.
To each data set, we add a Gaussian noise term with a standard
deviation of $\sigma_0 = 0.1$.

From the synthetic data sets we derive estimates of the known input standard
deviation, $\sigma_0$, constructing correlated $\beta$ samples using various
orders of approximation and the shifting procedure.
For each combination of period and order of approximation, we repeat the
experiment 200~times and record the resulting estimates of the standard
deviation, $s_E$, its estimated standard error, $\sigma_{s_E}$, and the
relative deviation from the known value $d=(s_E -
\sigma_0)\,\sigma_{s_E}^{-1}$.
In Table~\ref{tab:sine} we show the thus derived expected relative deviation of
the estimate from the true value for orders of approximation $0 \le N \le 5$.
Lower orders of approximation suffice in
the case of well sampled curves for which $P\Delta
t^{-1}$ is large. As the sampling of the oscillation becomes more sparse, the
required order of approximation rises. These effects are clearly seen in
Table~\ref{tab:sine}.

\begin{table}
\caption{Expected relative deviation of the estimated standard deviation from
the input (see Sect.~\ref{sec:sineApp}). Black numbers indicate consistency with the input to within two standard errors
and red numbers indicate larger expected deviations.
\label{tab:sine}}
\begin{tabular}{r r r r r r r} \hline\hline
$P\Delta t^{-1}$ & 0 & 1 & 2 & 3 & 4 & 5 \\ \hline
200.0 &  1.53 &  0.05 & -0.00 & -0.18 &  0.04 &  0.00\\ 
100.0 & {\color{red} $ 5.52$} & -0.03 & -0.01 &  0.04 & -0.00 & -0.08\\ 
 50.0 & {\color{red} $ 13.74$} &  0.20 & -0.09 & -0.08 &  0.00 & -0.06\\ 
 25.0 & {\color{red} $ 22.95$} &  1.98 & -0.06 & -0.03 & -0.13 & -0.03\\ 
 12.5 & {\color{red} $ 29.28$} & {\color{red} $ 13.66$} & {\color{red} $ 2.01$} &  0.20 & -0.08 & -0.06\\ 
 10.0 & {\color{red} $ 30.65$} & {\color{red} $ 18.81$} & {\color{red} $ 5.84$} &  0.81 & -0.07 &  0.04\\ 
  9.0 & {\color{red} $ 31.20$} & {\color{red} $ 20.91$} & {\color{red} $ 8.72$} &  1.67 &  0.20 & -0.06\\ 
  8.0 & {\color{red} $ 31.75$} & {\color{red} $ 22.94$} & {\color{red} $ 12.42$} & {\color{red} $ 3.68$} &  0.55 &  0.00\\ 
  7.0 & {\color{red} $ 32.30$} & {\color{red} $ 24.84$} & {\color{red} $ 16.58$} & {\color{red} $ 7.68$} & {\color{red} $ 2.19$} &  0.51\\ 
  6.0 & {\color{red} $ 32.85$} & {\color{red} $ 26.56$} & {\color{red} $ 20.51$} & {\color{red} $ 13.52$} & {\color{red} $ 6.62$} & {\color{red} $ 2.36$}\\ 
\hline
\end{tabular}
\end{table}

\subsubsection{A pathological case}
\label{sec:pathoCase}
A pathological case in the sense of convergence of the \bsp\ is the data set
$y_i = (-1)^i$, which could, for example, have been obtained by sampling only the
minima and maxima of an underlying oscillation. In this example, we 
assume a noise-free data set.
For any odd jump parameter, all realizations of $\beta$
will have the same value
\begin{equation}
  \beta_{m,N} = \frac{\sum_{k=0}^{N+1} (-1)^{k+i_0} a_k
  }{\sqrt{\binom{2N+2}{N+1}}} = \pm \frac{\sum_{k=0}^{N+1} \binom{N+1}{k}}{\sqrt{\binom{2N+2}{N+1}}}
  = \pm \frac{2^{N+1}}{\sqrt{\binom{2N+2}{N+1}}} \; ,
\end{equation}
where the sign depends on whether the starting index, $i_0$, of the subsample is
even or odd.

When the order of approximation is increased by one, the value of the
corresponding realizations of $\beta$ grows
\begin{equation}
  \frac{\beta_{m, N+1}}{\beta_{m,N}} = \sqrt{\frac{2N+4}{2N+3}} > 1 \; .
\end{equation}
If the number of data points is sufficiently large to apply arbitrary orders of
approximation, $N$, it can be shown that the variance estimate, $s^2_E$,
increases without bounds. Specifically, $s^2_E$ grows as
\begin{equation}
  s_E^2 = \frac{1}{n} \sum_{i=1}^n \beta_i^2 = \beta_0^2 \sim \sqrt{N} \; .
\end{equation}
Although the case here considered is somewhat contrived, it clearly demonstrates
that, first, there may be cases when the \bso\ procedure does not
converge to the true value of the noise for \emph{any} order of approximation and,
second, that estimates obtained from consecutive orders of approximation may be
arbitrarily close nonetheless. However, the fact that the estimates
continuously grow for increasing orders of approximation clearly indicates that
no improvement is achieved and the underlying approximations may be violated.
In this specific case, we note that for even jump parameter all realizations of
$\beta$ are zero and the correct result would be derived, which is due to the
special construction of the problem considered here.

\subsection{Application to high-resolution spectra}
\label{sec:AppliSpec}
\begin{figure}
    \includegraphics[width=0.48\textwidth]{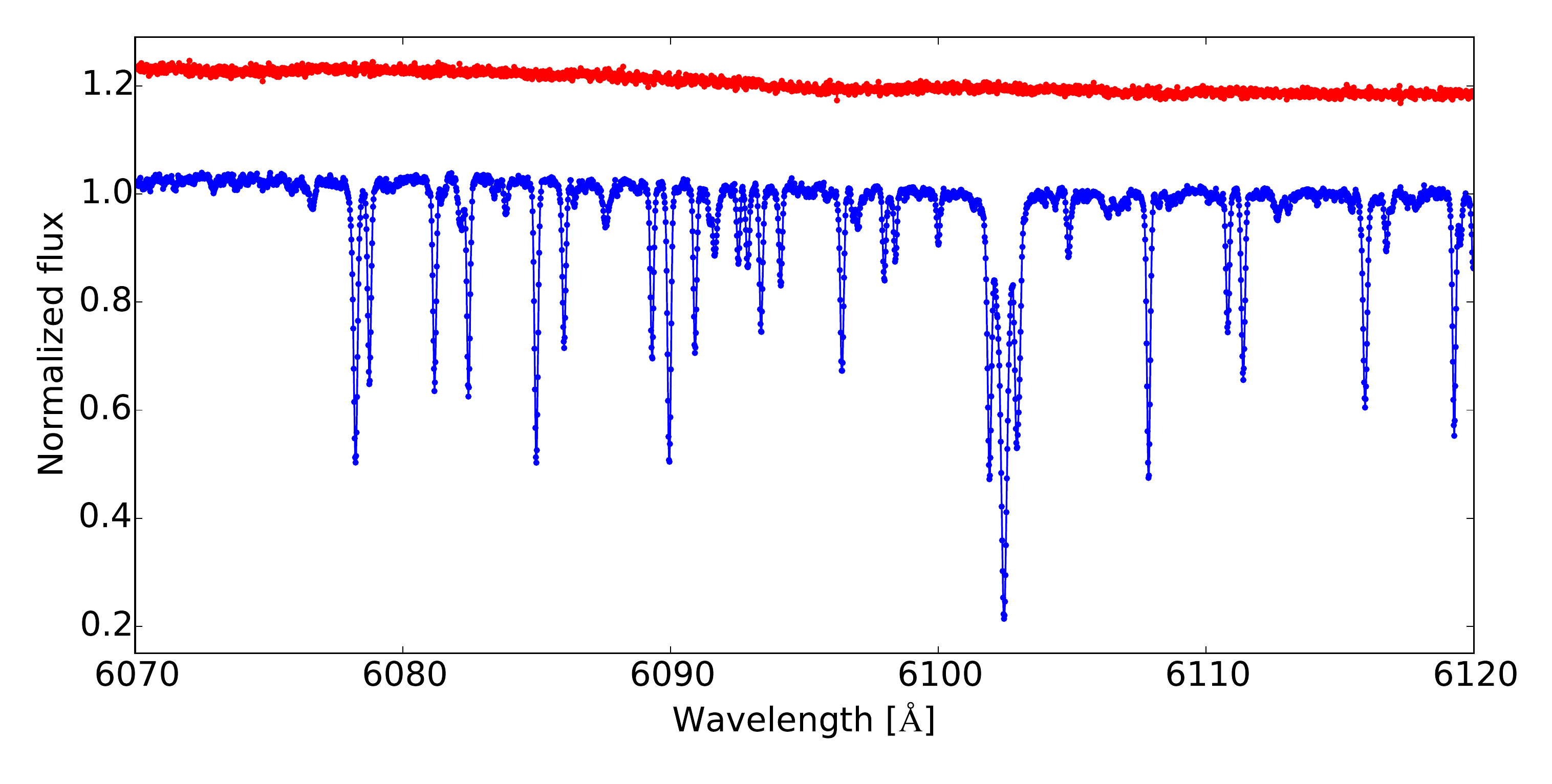}
  \caption{Normalized spectrum of \hr\ (red, shifted by +0.2) and \hd\ (blue).
  \label{fig:specs}}
\end{figure}

We obtain estimates of the noise in high-resolution echelle spectra of \hr\
and \hd\ obtained with the UVES spectrograph\footnote{program
ID: 089.D-0701(A)}. In our analysis, we adopted
the $6070-6120$~\AA\ range. The corresponding spectra are shown
in Fig.~\ref{fig:specs}.
The data reduction is based on the UVES pipeline and is
described in \citet{Czesla2015}.

\begin{figure}
    \includegraphics[width=0.48\textwidth]{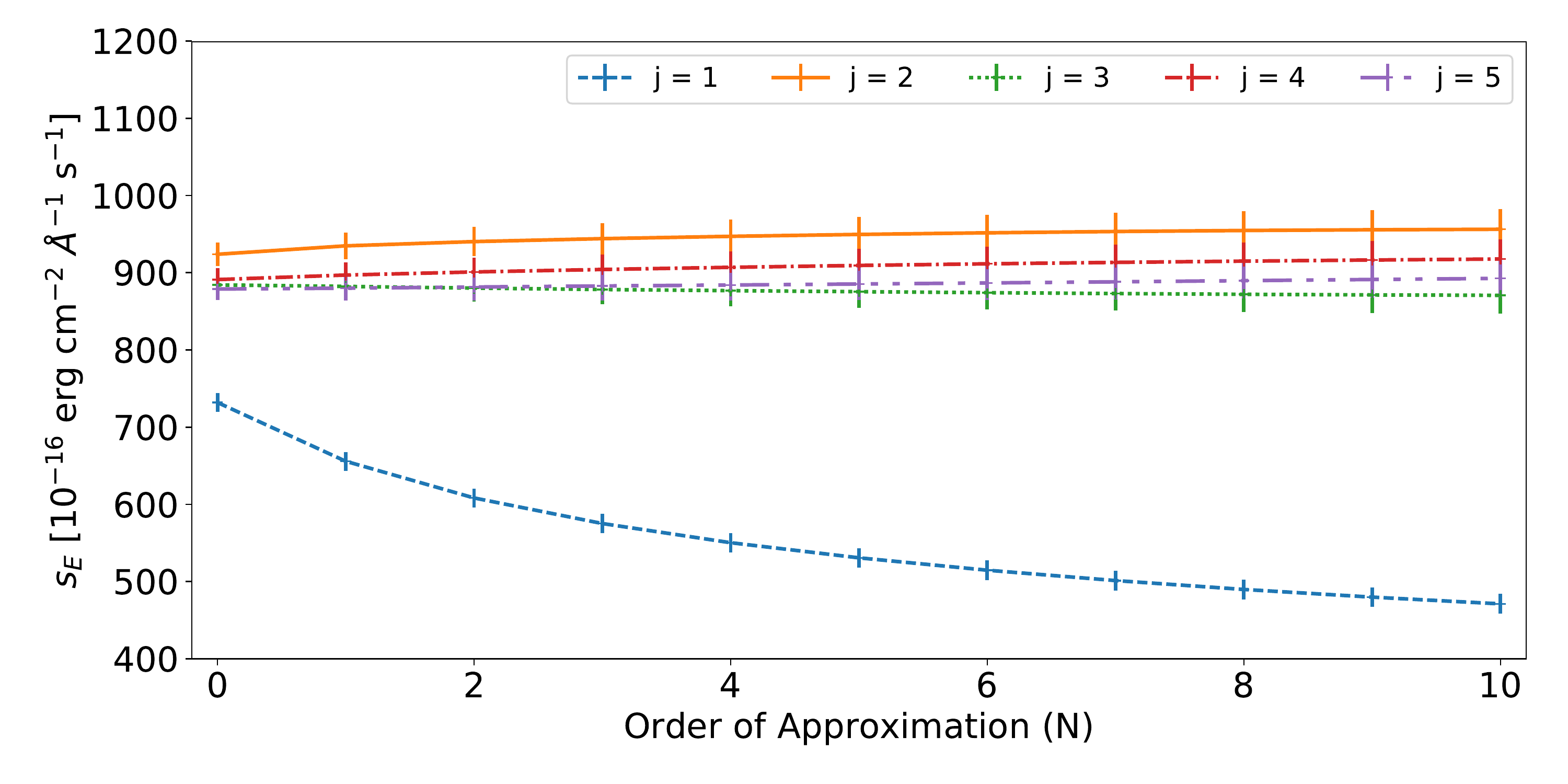}
  \caption{Noise estimate as a function of order of
  approximation, $N$, and jump parameter, $j$, for \hr.
  \label{fig:bsHR}}
\end{figure}

First, we estimate the noise in the spectrum of the fast-rotating B3V
star \hr, which shows no narrow spectral features and served as a telluric
standard.
In Fig.~\ref{fig:bsHR}, we show the estimated standard deviation of the
noise, obtained for different orders of
approximation and jump parameters, using the MV estimator; the $\beta$ sample
was constructed using the shifting procedure.
The estimates
obtained for a jump parameter of one clearly converge to a lower
value for increasing orders of approximation than those for larger jump
parameters.
As the data do not show strong spectral lines, we attribute
this behavior to correlation in the noise of adjacent data points, so
that the assumption of mutually independent realizations of noise
is violated. 
For larger jump parameters the estimates hardly vary as a function of the
order of approximation,
showing the high degree of correlation demonstrated in Sect.~\ref{sec:estiCorr}.  
Valid estimates are thus only obtained for jump parameters larger than one; all
such estimates for \hr\ considered in our analysis shown in
Fig.~\ref{fig:bsHR} are essentially consistent.
The picture obtained by using the robust
estimator is very similar.

Combining the estimate of $(884\pm 14)\times
10^{-16}$~\mbox{erg\,cm$^{-2}$\,$\AA^{-1}$\,s$^{-1}$} obtained for the zeroth
order of approximation and a jump parameter of three with the median flux
density in the studied spectral range, we obtain an estimate of $209\pm 4$ for the
signal-to-noise ratio. The estimates based on the MV and robust estimators
provide consistent results and, naturally, also the \ds\
estimate is compatible with this value.
While this result appears to be consistent with an estimated
scatter of $993\times 10^{-16}$~\mbox{erg\,cm$^{-2}$\,$\AA^{-1}$\,s$^{-1}$} in the residuals after
a fourth order polynomial approximation of the entire considered range, the mean pipeline
estimate of about $1500$~\mbox{erg\,cm$^{-2}$\,$\AA^{-1}$\,s$^{-1}$} appears to
be on the conservative side. As a cross-check of our estimate, we added
uncorrelated noise with a standard deviation of
$500$~\mbox{erg\,cm$^{-2}$\,$\AA^{-1}$\,s$^{-1}$} to the data. This yielded 
a noise estimate of $1017\pm 14$~\mbox{erg\,cm$^{-2}$\,$\AA^{-1}$\,s$^{-1}$} for
the same order of approximation and jump parameter, which is consistent with the
sum of variances.

\begin{figure}
    \includegraphics[width=0.48\textwidth]{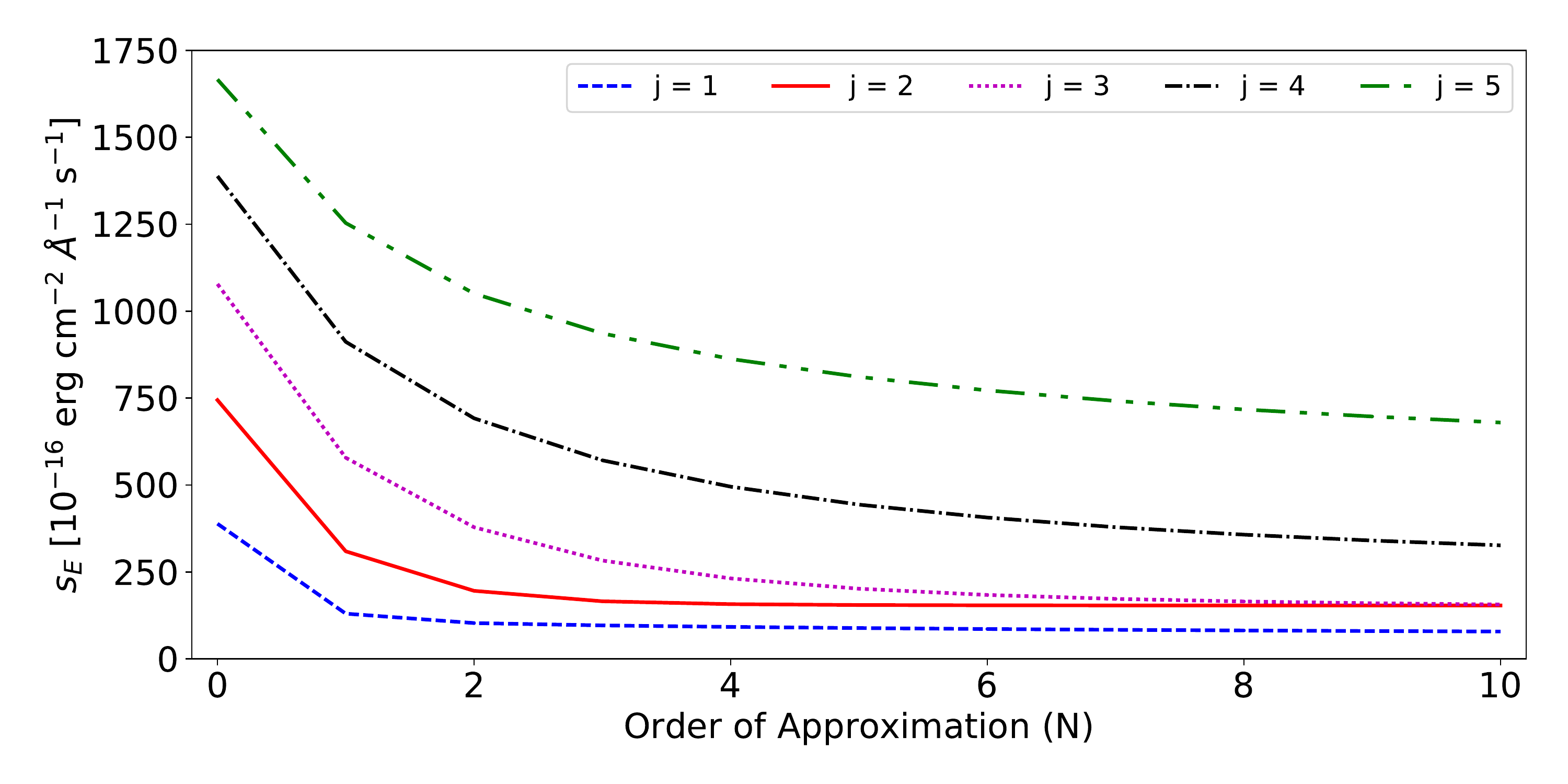}
  \includegraphics[width=0.48\textwidth]{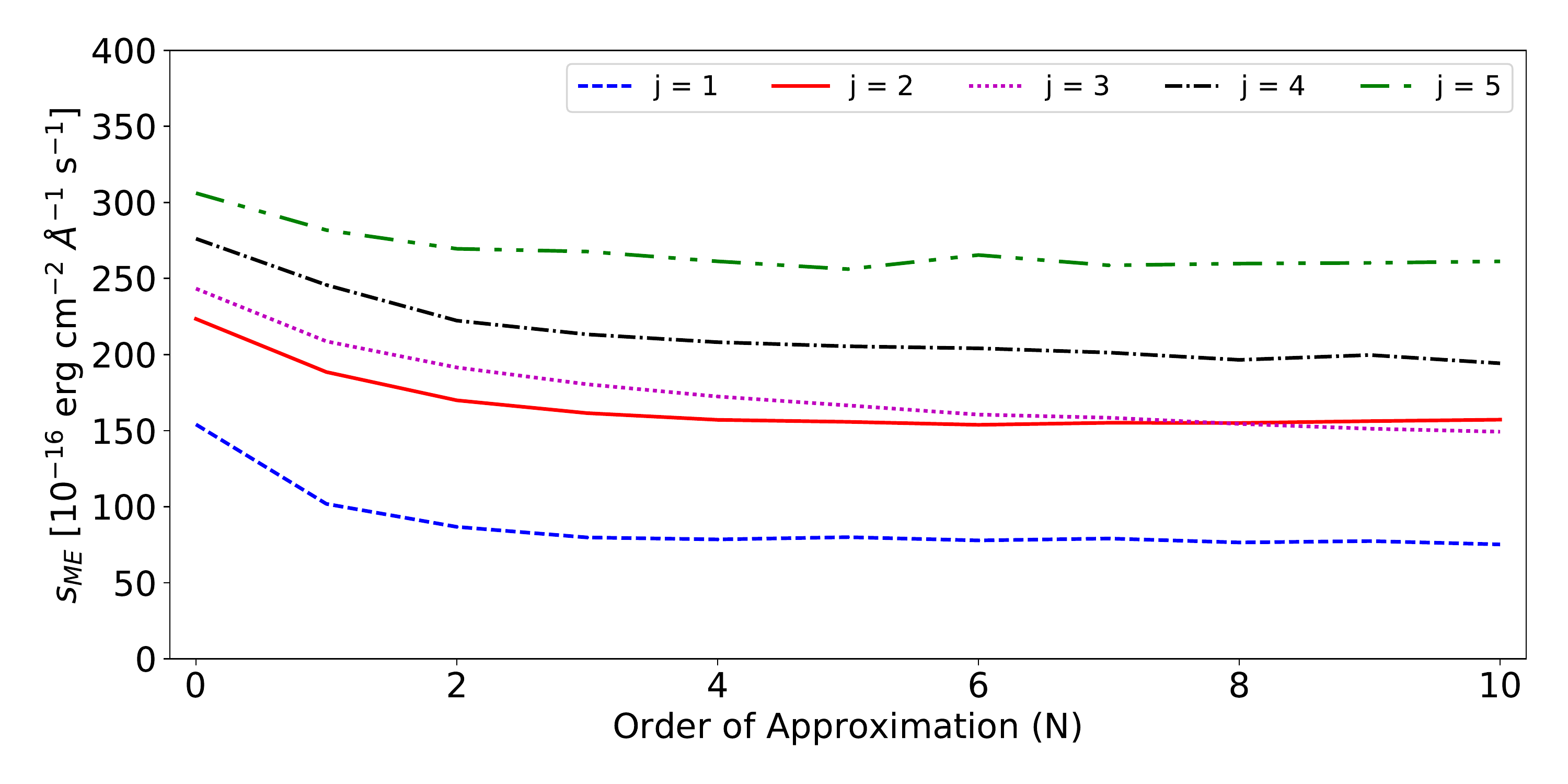}
  \caption{Noise estimates for the spectrum of \hd\ as a function
  of order of approximation, $N$, and jump parameter, $j$, for efficient (top)
  and robust (bottom) estimator.
  \label{fig:bsHD}}
\end{figure}

In Fig.~\ref{fig:bsHD}, we show the noise estimates obtained from the spectrum
of the K-type star \hd\ using the MV estimator, $\hat{s}_E$, and the robust
estimator, $\hat{s}_{\rm ME}$.
For a jump parameter of one, the estimates again converge to a lower value than for
larger jump parameters, which we attribute
to correlation. The series of estimates obtained with the robust estimator show a
much better convergence behavior toward higher orders of approximation.
In particular, lower orders of approximation tend to yield significantly smaller
robust estimates; we emphasize the difference in scale in the top and
bottom panel.

While for jump parameters of two and three, both the MV and robust
estimates converge toward a value of
about $150$~\mbox{erg\,cm$^{-2}$\,$\AA^{-1}$\,s$^{-1}$}, larger jump parameters
systematically yield larger estimates.
We attribute this behavior to the structure seen in the spectrum,
that is, the spectral lines. The larger the jump parameter, the longer the sections
of the spectrum, which have to be approximated by a polynomial of the same
degree. 

To obtain reliable error estimates, a jump parameter of two or three and an
order of approximation larger than about four appear necessary in this case.
Adopting the robust estimate $s_{\rm ME}(N=5, j=2)$ of $(156 \pm
4) \times 10^{-16}$~\mbox{erg\,cm$^{-2}$\,$\AA^{-1}$\,s$^{-1}$},
we obtain a signal-to-noise estimate of $205\pm 8$
in the considered spectral range. We approximate the variance of the robust
estimator by scaling that of the efficient estimator with a factor of $2.7$, in
accordance with their asymptotic efficiencies.
Again, the MV and robust estimates are consistent. Using the \ds\ in
its original form (i.e., $N=1$ and $j=2$) yields a robust noise estimate about
$20$\% larger than the value adopted here.

From the above, it is clear that reliable noise estimates need some scrutiny.
In the case at hand, the availability of the spectrum of \hr\ is of great help because
it allows \textbf{us} to study the noise properties in a reference data set
largely free of any underlying variability, which immediately reveals the peculiarity of the results
for unity jump parameter.
Nonetheless, also given the data of \hd\ alone, this would have been recognized,
and a reliable estimate could have been derived.

\subsection{Application to CoRoT data}
\label{sec:corotExample}
Next,
we obtain estimates of the noise contribution in a sample of CoRoT light
curves using the \bsp.
Specifically, we analyze all 1640 short-cadence light curve from the fourth long run, targeting
the galactic anticenter \citep[LRa04, e.g.,][]{Auvergne2009}. Because the number
of data points per light curve is large ($\approx 180\,000$), we construct
independent $\beta$ samples. Estimates obtained using an order of
approximation, $N$ and jump parameter, $j$, are denoted by $\bs{N}{j}$.

To
determine the required order of approximation and the jump parameter, we apply the following procedure:
First, we use an order of approximation of zero and a jump parameter of
one and obtain the estimates $\bs{N}{j}$, $\bs{N+1}{j}$, and
$\bs{N+1}{j+1}$ using the robust estimator, taking into account
irregular sampling. Second, we check whether the three estimates are consistent,
which we define to be the case when their three sigma confidence intervals
overlap. Again, we approximate the variance of the robust estimator by scaling
that of the MV estimator in accordance with their
asymptotic efficiencies. If $\bs{N}{j}$ and
$\bs{N+1}{j}$ are inconsistent, we increase the order of approximation by one
and restart the procedure. If $\bs{N}{j}$ and $\bs{N+1}{j+1}$ are
inconsistent, we increase both the order of approximation and the jump
parameter by one and restart the procedure. If $\bs{N}{j}$ is consistent with
the other two estimates, we stop the iteration.

\begin{figure}
  \includegraphics[width=0.49\textwidth]{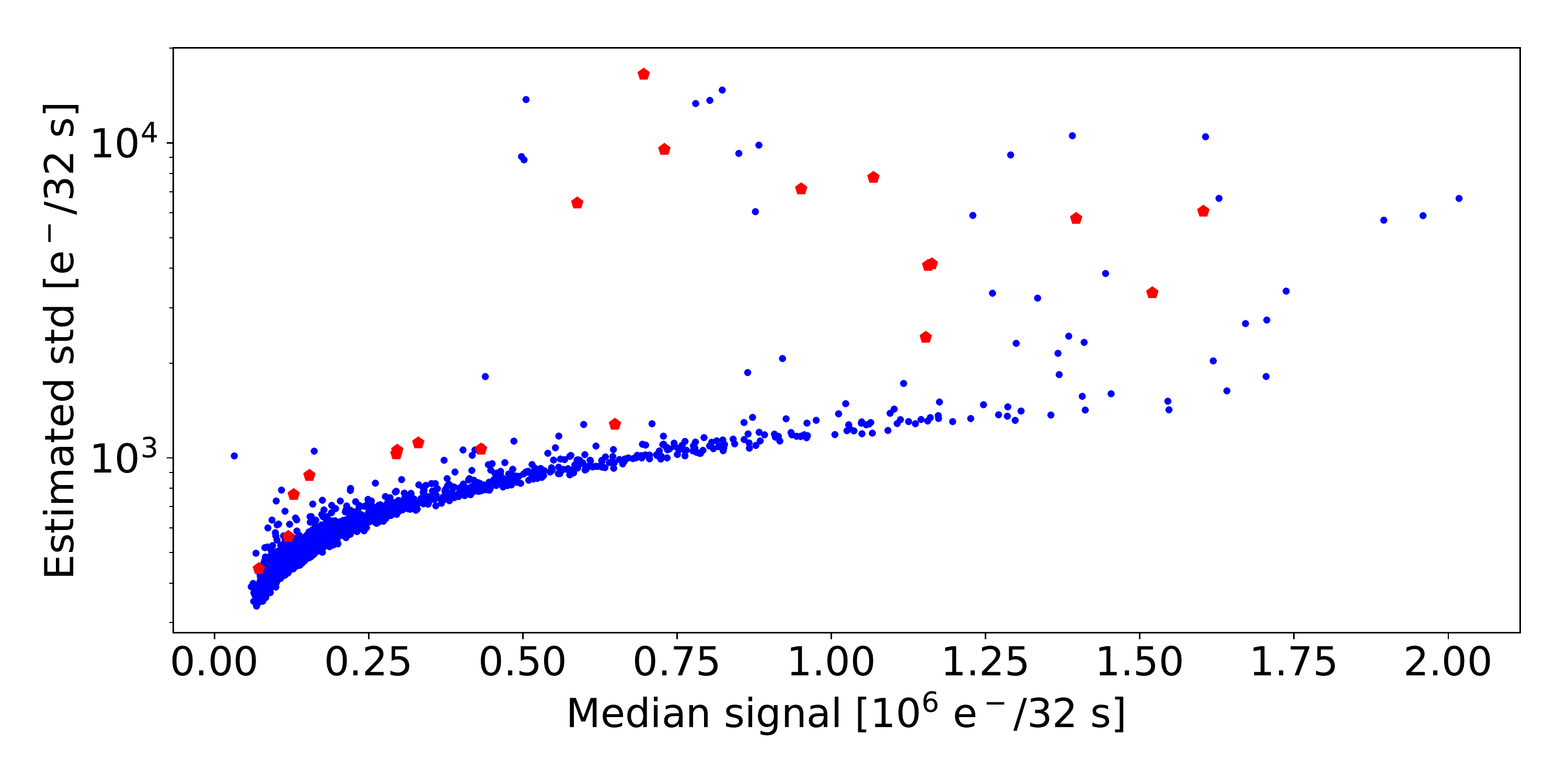}
  \caption{Estimated standard deviation of noise as a function of median signal
  in the LRa04 short-cadence light curves obtained by CoRoT (blue dots). Red
  points indicate light curves for which our iterative procedure yielded an order of
  approximation larger than two.
  \label{fig:Corot}}
\end{figure}

For 1499 out of 1640 light curves, the zeroth order of approximation remained
sufficient according to our criterion, 33 required the first, and 88 the second
order; in 20 cases a higher order was required.
In Fig.~\ref{fig:Corot} we show the resulting
noise estimate as a function of the median flux in the light curve.
The distribution shows that the noise contribution qualitatively follows a
square-root relation with respect to the median flux,
which is indicative of a dominant Poisson noise contribution.
The results in Fig.~\ref{fig:Corot} have been derived
allowing for arbitrary sampling, however, the majority of
robust estimates obtained assuming regular sampling are identical.
If there is a difference, estimates based on equidistant sampling tend to be
larger by typically less than a percent.
Compared to the estimates obtained using the MV estimator, the
robust estimates tend to be lower by a few percent for the order of
approximation and jump parameter at which the iterative procedure halted.

\begin{figure}
  \includegraphics[width=0.49\textwidth]{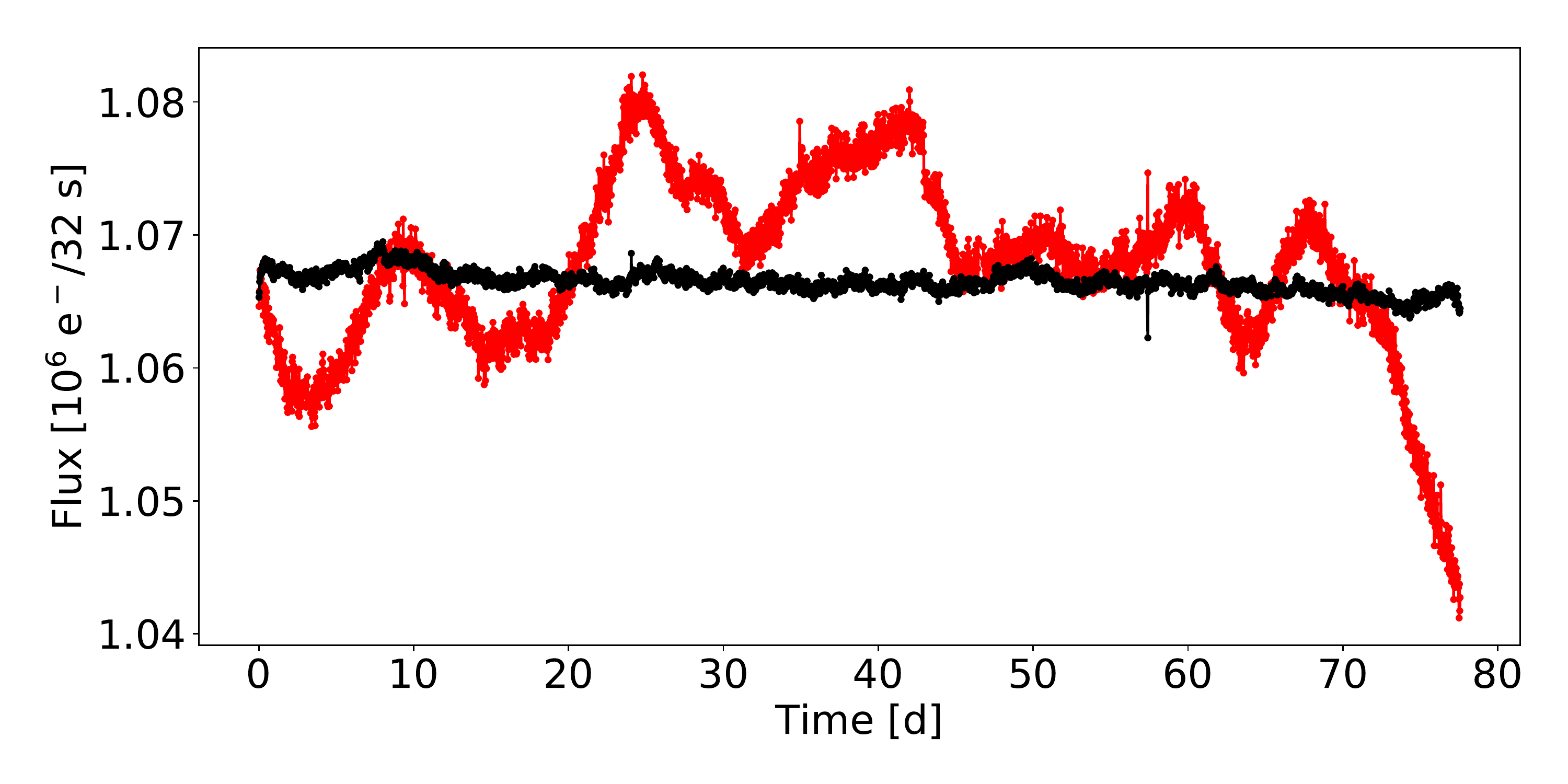}
  \includegraphics[width=0.49\textwidth]{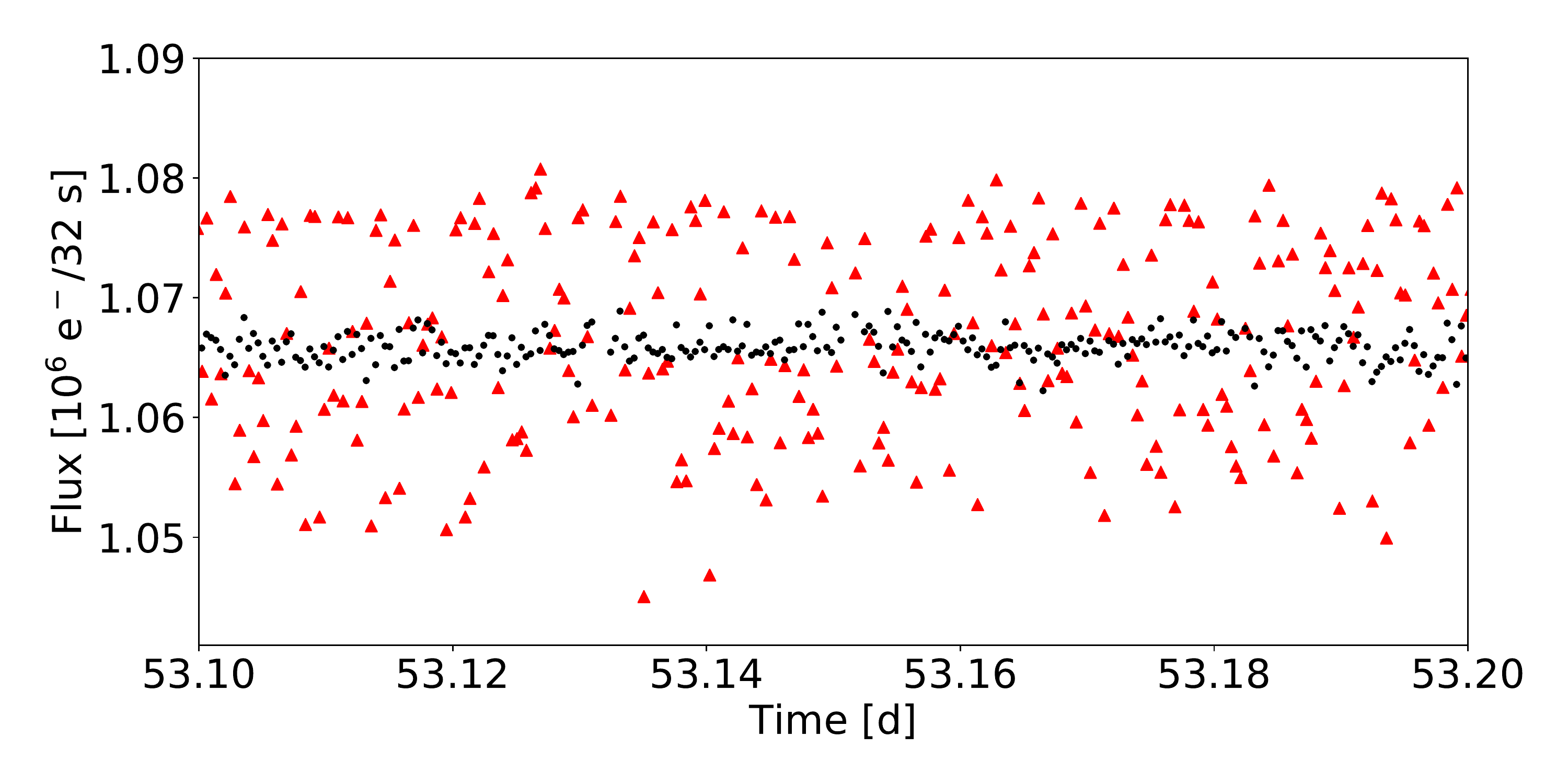}
  \caption{Light curves of the stars with CoRoT IDs 605144111
  (black) and 605088599 (red). Top panel: Entire available light
  curve with a temporal binning of $30$~min. Bottom panel: Excerpt of the light
  curves with original binning of $32$~s (black dots for CoRoT IDs 605144111 and
  red triangles for 605088599).
  \label{fig:C2LC}}
\end{figure}

A number of light
curves apparently show noise levels significantly above the square-root
relation outlined by the majority of noise estimates.
In Fig.~\ref{fig:Corot}, we indicate those light
curves for which our iterative procedure halted at an order of approximation
larger than two. While the corresponding light curves tend to be those with
higher noise levels, there is no unique mapping.
We checked our result
using, first, a comparison with the noise level estimated using the standard
derivation of the residuals obtained from a second-order polynomial fit to the
first $200$ data points in each light curve. The noise estimates obtained with
this method are on average $6$\% higher than those derived using the \bsp\ with
a number of estimates off by a factor of a few. Yet,
the result confirms the reality of the outliers.
Second, a visual inspection confirmed a comparatively large scatter
in the light curves, consistent with the obtained estimates.
Light curves showing increased noise levels were also
reported on by \citet{Aigrain2009}, who studied the noise in CoRoT light curves
on the transit timescale.

As an example,
Fig.~\ref{fig:C2LC} shows the light curves of the
stars with CoRoT IDs 605144111 and 605088599,
which both show similar median flux levels of $1.06\times 10^6$~\mbox{e$^- (32
\mbox{s})^{-1}$}. While the estimated scatter in the first is consistent with
the square-root relation, the second is a factor of $6.8$
larger. For this star, Fig.~\ref{fig:C2LC} shows both a larger level of
(apparently) stochastic variability between individual data points (bottom
panel) and a higher degree of overall variability in the $80$~d long light curve
(top panel), which remains, however, irrelevant in the noise estimation.
In fact, the light curve of CoRoT~605088599 is among the few for which
our procedure halted only at an order of approximation of seven with a jump
parameter of six.

The behavior of the outlier light curves may be due to an additional source of
random noise or a highly variable stochastic component in the light curve both
either of instrumental or
physical origin. The \bsp\ provides an
estimate of the standard deviation of the stochastic contribution in the data as
specified in Sect.~\ref{sec:Method}; it neither distinguishes
between individual sources of that term nor does it tell us anything about its
origin in itself. It is also conceivable that actual convergence has not been
achieved in some cases (see Sect.~\ref{sec:pathoCase}).
While an in-depth analysis of the peculiar light curves forming the class of
outliers is beyond the scope of this discussion, the presented procedure proved
highly robust in estimating the amplitude of noise in a large sample of light
curves with a wide range of morphologies.

\section{Setting up the \bsp}
\label{sec:Setup}

The \bsp\ as presented here is a concept, requiring a number of choices to
be made prior to application. Specifically, the order of approximation and the
jump parameter have to be selected, a strategy to construct the $\beta$
sample has to be chosen, and an estimator to determine the standard deviation
(or variance) of the $\beta$ sample needs to be specified. A number of
options, but certainly not all, have been discussed above.

Clearly, the optimal choice depends on the data set at hand, and our decision
may be guided by the information we have about the data, such as a typical scale
of variation compared to the sampling cadence.
While it can be shown under which conditions the $\beta\sigma$ (or \ds)
estimates are reliable, we do not see a possibility to prove that the method yielded the
correct result based on the result itself.

In fact, any set of $N+2$ measurements may also be described by a
polynomial of degree $N+1$ with no random noise contribution at all.
If $g(t)$ is a function showing $N$ or more extrema in the range
covered by the $N+2$ data points such as a higher-order polynomial or fast
oscillation, the proposed estimator will generally be biased because an
$N^{\rm th}$ order polynomial approximation remains no longer appropriate.
Nonetheless, useful strategies and
recipes can be defined to strengthen the confidence in the outcome.

\subsection{Choosing the order of approximation and the jump parameter}
\label{sec:OrderOfApp}
The required order of approximation is mainly determined by the
degree of variability seen in the data with respect to the sampling rate. Slowly
varying (or better sampled) data sets require lower orders of approximation.  
The larger the jump parameter is chosen, the longer become the
sections of the data, which have to be modeled using the same order of
approximation. If correlation in the noise distribution is known, it is
advisable to opt for a jump parameter larger than the correlation scale.

Based on our experience, we propose to obtain at least two
estimates with consecutive orders of approximation and accept the values only
if both estimates are consistent.
If the conditions for the method to be applicable hold, the estimates
obtained using consecutive orders of estimates are expected to yield
statistically indistinguishable results (Sect.~\ref{sec:estiCorr}). 
If this is not the case, a higher order of approximation may be
required to account for the intrinsic variation in the data, and the order should be
increased by one.
Unless other prior information is available, we suggest to start with the lowest
orders of approximation (zero and one).
At any rate, the consistency of estimates obtained with different orders of
approximation is not a sufficient condition for a valid estimate
(Sect.~\ref{sec:pathoCase}).

Additionally, we suggest to carry out a cross-check of the result using a
a larger jump parameter to exclude effects, for example, attributable to correlation
between the noise in adjacent data points. Such an effect was observed in our
study of the high-resolution spectra (Sect.~\ref{sec:AppliSpec}) and was also
discussed by \citet{Stoehr2008}. No such effect appears to be present in the
CoRoT data (Sect.~\ref{sec:corotExample}).

\subsection{Selecting an estimator}

For well-behaved, Gaussian $\beta$ samples, the estimator used to obtain the
standard deviation remains of little practical significance. In working with
real data, it appears that the main choice is between robust and
non-robust techniques.
While the robust estimators are typically of smaller efficiency (for Gaussian
samples), we attribute
the main problems in our applications to outliers and non-Gaussian samples.
Our results suggest that robust estimates are more reliable.
We caution that in the case of small $\beta$ samples, estimation biases may
become an issue to be kept in mind.

\subsection{Construction of the $\beta$ sample}
We discussed procedures to construct both $\beta$ samples with independent and
correlated elements. We found the estimates derived from the correlated
$\beta$ samples, constructed using the shifting technique, superior in many
applications
because
they preserve a larger fraction of the information contained in the data. The
uncorrelated $\beta$ samples are smaller in size and give rise to less
precise noise estimates. Nonetheless, independent samples may still be useful, for example, if
estimation biases due to correlation shall be excluded or the number of data
points is large (Sect.~\ref{sec:corotExample}).

\section{Conclusion}
\label{sec:Conclusion}

The \bsp\ presented here is a versatile technique to estimate the amplitude of
noise in discrete data sets, which can be proved to work for sufficiently
well-sampled data sets. It is related to
numerical differentiation and differencing as regularly applied in time series analyses
\citep{Shumway}. Also polynomial approximations have been used for a long
time, for example, in the context of the Savitzky-Golay filter \citep{Savitky}.
However, we are not aware of an application of the presented procedure in
astronomy other than in the form of the
\dsa\ by \citet{Stoehr2008}; in fact, the \bsp\ \emph{is} the \dsa\ for a
specific choice of parameters.

We provide an analysis of the statistical properties of the resulting noise
estimates, depending on the chosen parameters for the \bsp\ and
applied it to synthetic data,
high-resolution spectra, and a large sample of CoRoT light curves.
While the conditions for the procedure to work can clearly be spelled out,
a difficulty in the application arises from the
fact that we cannot show the validity of the noise estimate by
application of the procedure itself.

In our test applications, we address this problem by comparison of estimates
obtained from a number of
$\beta$ samples, constructed using different orders of
approximation and jump parameters. Noise estimates are only accepted
if they are consistent. Unless other external
information is available about the data, this is also the recipe we suggest to
be used in the general case of application. In our test cases,
robust estimation proved to be advantageous in determining the
amplitude of noise.
While we agree with \citet{Stoehr2008} that the settings of the \dsa\ (i.e.,
$N=1$ and $j=2$) are reliable in many cases, we cannot
generally conclude that this will be the case for any data set. Therefore,
we suggest to verify the result by comparing it with that obtained using the
next higher order of approximation and eventually, also a larger (or smaller)
jump parameter.

Along with this paper, we provide a \texttt{Python} implementation of the
\bsp, open for use and modification by all interested
parties. The \bsp\ is highly versatile and once the properties of the data
are approximately known, it can be applied to derive the amplitude of noise in
large samples without further interference.

\begin{acknowledgements}
We thank the anonymous referee for knowledgeable comments, which helped to
improve our paper.
\end{acknowledgements}

\bibliographystyle{aa}
\bibliography{doc}

\end{document}